\documentclass[11pt]{article}

\RequirePackage[OT1]{fontenc} \RequirePackage{amsthm,amsmath}
\usepackage[round]{natbib}
\RequirePackage[colorlinks,citecolor=blue,urlcolor=blue]{hyperref}

\usepackage{graphics,epsfig,epsf,psfrag}
\usepackage{url}
\usepackage{color, appendix}
\usepackage{graphicx}
\usepackage{amsmath, amsfonts, amsthm,amssymb}
\usepackage{epstopdf}
\usepackage{hyperref}
\usepackage{enumerate, framed}
\usepackage{graphicx,lscape,rotate}
\usepackage{algorithmic, algorithm}

\let\BLS=\baselinestretch
\makeatletter
\newcommand{\singlespacing}{\let\CS=\@currsize\renewcommand{\baselinestretch}{1}\small\CS}
\newcommand{\doublespacing}{\let\CS=\@currsize\renewcommand{\baselinestretch}{1.5}\small\CS}
\newcommand{\normalspacing}{\let\CS=\@currsize\renewcommand{\baselinestretch}{\BLS}\small\CS}
\makeatother

\numberwithin{equation}{section}
\theoremstyle{plain}
\newtheorem{thm}{Theorem}[section]
\newtheorem*{remark}{Remark}

\newtheorem*{definition}{Definition}
\theoremstyle{remark}

\def\keywords{\vspace{1.5em}
{\noindent \textbf{Key Words}:\,\,\relax%
}}

\newcommand{\bfbeta}{\mbox{\boldmath $\beta$}}

\newcommand{\bfSigma}{\mbox{\boldmath $\Sigma$}}

\newcommand{\bfgamma}{\mbox{\boldmath $\gamma$}}

\newcommand{\X}{\mathbf{X}}

\newcommand{\x}{\mathbf{x}}

\newcommand{\y}{\mathbf{y}}

\newcommand{\diag}{\mbox{diag}}
\newcommand{\sgn}{\mbox{sgn}}
\newcommand{\tr}{\mbox{tr}}

\newcount\colveccount
\newcommand*\colvec[1]{
        \global\colveccount#1
        \begin{pmatrix}
        \colvecnext
}
\def\colvecnext#1{
        #1
        \global\advance\colveccount-1
        \ifnum\colveccount>0
                \\
                \expandafter\colvecnext
        \else
                \end{pmatrix}
        \fi
}

\makeatletter
\newcommand{\Spvek}[2][r]{%
  \gdef\@VORNE{1}
  \left(\hskip-\arraycolsep%
    \begin{array}{#1}\vekSp@lten{#2}\end{array}%
  \hskip-\arraycolsep\right)}

\def\vekSp@lten#1{\xvekSp@lten#1;vekL@stLine;}
\def\vekL@stLine{vekL@stLine}
\def\xvekSp@lten#1;{\def\temp{#1}%
  \ifx\temp\vekL@stLine
  \else
    \ifnum\@VORNE=1\gdef\@VORNE{0}
    \else\@arraycr\fi%
    #1%
    \expandafter\xvekSp@lten
  \fi}
\makeatother

\begin{document}


\title{Sparse Estimation of Generalized Linear
Models (GLM) \\ via Approximated Information Criteria}

\author{\textbf{Xiaogang Su}\footnote{Email: \texttt{xsu@utep.edu}} \\
Department of Mathematical Sciences, University of Texas, El Paso, TX 79968  \\
\and \textbf{Juanjuan Fan}, \textbf{Richard A.~Levine} \\
Department of Mathematics \& Statistics, San Diego State University, CA 92182 \\
\and \textbf{Martha E.~Nunn} \\
Department of Periodontology, Creighton University, Omaha, NE 68178 \\
\and \textbf{and Chih-Ling Tsai} \\
Graduate School of Management, University of California, Davis, CA
95616}


\date{}  

\maketitle

\renewcommand{\abstractname}{\large Abstract}
\begin{abstract}
We propose a new sparse estimation method, termed MIC (Minimum
approximated Information Criterion), for generalized linear models
(GLM) in fixed dimensions. What is essentially involved in MIC is
the approximation of the $\ell_0$-norm with a continuous unit dent
function. Besides, a reparameterization step is devised to enforce
sparsity in parameter estimates while maintaining the smoothness
of the objective function. MIC yields superior performance in
sparse estimation by optimizing the approximated information
criterion without reducing the search space and is computationally
advantageous since no selection of tuning parameters is required.
Moreover, the reparameterization tactic leads to valid
significance testing results that are free of post-selection
inference. We explore the asymptotic properties of MIC and
illustrate its usage with both simulated experiments and empirical
examples.
\end{abstract}


\section{Introduction}

Suppose that data $\mathcal{L} := \{(y_i, \mathbf{x}_i): i=1,
\ldots, n\}$ consist of $n$ i.i.d.~copies of $\{y, \mathbf{x}\}$,
where $y$ is the response variable and $\mathbf{x} = (x_1, \ldots,
x_p)^T \in \mathbb{R}^p$ is the predictor vector. WLOG, we assume
that the $x_{ij}$'s are standardized throughout the paper.
Consider the regression models that link the mean response $y$ and
covariates $\x$ through its linear predictor $\x^T\bfbeta$ with
$\bfbeta=(\beta_1,\ldots,\beta_p)^T$, e.g., generalized linear
models (GLM; \citeauthor{McCullagh.Nelder.1989},
\citeyear{McCullagh.Nelder.1989}). Concerning variable selection,
the true $\bfbeta$ is often sparse in the sense that some of its
components are zeros. To this end, we assume that either there is
no nuisance parameter involved or the nuisance parameters and
$\bfbeta$ are orthogonal \citep{Cox:2008}. Hence we simply denote
the log-likelihood function as $L(\bfbeta) = \sum_{i=1}^n \log
f(y_i; \mathbf{x}_i, \bfbeta).$

A classical variable selection procedure is the best subset
selection (BSS), and two commonly-used information criteria are
AIC \citep{Akaike.1974} and BIC \citep{Schwarz.1978}. BSS can be
formulated as
\begin{equation}
\min_{\bfbeta \, \in \, \Omega} ~~ - 2 \cdot L(\bfbeta) +
\lambda_0 \cdot \parallel \bfbeta \parallel_0, \label{best-subset}
\end{equation}
where the penalty parameter $\lambda_0$ is fixed as 2 in AIC or
$\ln(n)$ in BIC. We shall focus more on the use of BIC for its
superior empirical performance in variable selection widely
reported in the literature. In addition, the `$\ell_0$-norm'
$\parallel \bfbeta
\parallel_0 = \mbox{card}(\bfbeta) = \sum_{j=1}^p I( \beta_j \neq
0 )$ denotes the cardinality or the number of nonzero components
in $\bfbeta$ and the search space in (\ref{best-subset}) is the
entire parameter space $\Omega$ for $\bfbeta.$ Due to the discrete
nature of cardinality, $\Omega$ consists of $\bfbeta$ associated
with all possible $2^p$ sparsity structures. Optimization of
(\ref{best-subset}) proceeds in two steps: first maximize the
log-likelihood function for every known sparsity structure in
$\bfbeta$ and then compare the resulting information criteria $- 2
\cdot L(\widehat\bfbeta) + \lambda_0 \cdot
\parallel \widehat\bfbeta \parallel_0$ across all model
choices, where $\widehat\bfbeta$ denotes the maximum likelihood
estimator of $\bfbeta$ and hence $L(\widehat\bfbeta)$ corresponds
to the maximized log-likelihood function. While faster algorithms
\citep{Furnival.Wilson.1974} are available, solving
(\ref{best-subset}) is non-convex and NP-hard. As a result, the
best subset selection becomes infeasible when $p$ is moderately
large.

Both ridge regression \citep{Hoerl.Kennard.1970} and LASSO
\citep{Tibshirani.1996} were proposed as convex relaxations of
(\ref{best-subset}). Their general form is given by
\begin{equation}
\min_{\bfbeta} ~ - 2 \cdot L(\bfbeta) + \lambda \, \cdot
\parallel \bfbeta \parallel_r,
\label{bridge}
\end{equation}
where $\parallel \bfbeta \parallel_r = \sum_{j=1}^p |\beta_j|^r$
for some $r >0$. To assure convexity, $r \geq 1$ is typically
considered, and $r=1$ and $r=2$ result in LASSO and ridge
regression, respectively. While both methods provide a continuous
regularization process for ill-posed estimation problems, LASSO
enjoys the additional property of enforcing sparsity. Its
regularization path is shown to be piecewise-linear and can be
efficiently computed via either the homotopy algorithm
(\citeauthor{Osborne.etal.2000}, \citeyear{Osborne.etal.2000} and
and \citeauthor{Efron.etal.2004}, \citeyear{Efron.etal.2004}) or
the coordinate descent (\citeauthor{Fu.1998}, \citeyear{Fu.1998}
and \citeauthor{Friedman.etal.2010},
\citeyear{Friedman.etal.2010}). Since the proposal of LASSO, a
vast statistical literature has been devoted to the study of
$\ell_1$ regularization, and numerous variants have been developed
for enhancement and expansion. An up-to-date literature review
can be found in \citet{Zhang.2010}, \citet{Breheny.Huang.2011},
\citet{Shen.etal.2012} and references therein.

However, the convex relaxation methods with $\parallel \bfbeta
\parallel_r$ are mainly motivated by optimization theory; by no
means are they intended as an approximation of $\parallel \bfbeta
\parallel_0$ in (\ref{best-subset}). With the formulation
(\ref{bridge}), one would lose track of $\lambda_0$. As a result,
$\lambda$ in (\ref{bridge}) becomes a tuning parameter and its
choice has to be selected with extra efforts. The common practice
of regularization involves two steps as well: first compute the
whole regularization path, i.e., the solution of $\bfbeta$ for
every tuning parameter $\lambda \geq 0$, and then tune for the
best $\lambda^\star$ via some criterion such as cross validation
or BIC (see, e.g., \citeauthor{Wang.etal.2007},
\citeyear{Wang.etal.2007}). This practice amounts to first
reducing the search space from the $p$-dimensional $\Omega$ to a
one-dimensional curve (often termed as regularization path)
$\{\widehat{\bfbeta}(\lambda): \lambda \geq 0 \}$, and
subsequently selecting the best estimator
$\widehat{\bfbeta}(\lambda^\star)$. To have correct variable
selection, it is essential  that the true sparsity structure be
included in the much reduced search space, i.e., the
regularization path. However, this requirement cannot be
guaranteed for many existing $\ell_1$ regularization methods. In
particular, selection consistency of LASSO entails a strong
irrepresentable assumption \citep{Zhao.Yu.2006}. This has
motivated the proposals of non-convex penalties such as SCAD
\citep{Fan.Li.2001} and MCP \citep{Zhang.2010}. Another
statistically awkward issue with regularization is the selection
of the tuning parameter, which is conventional in optimization.
Selecting of the best tuning parameter $\hat{\lambda}$ is
computationally costly. Moreover, even though $\hat{\lambda}$ is
clearly a statistics, selection of the tuning parameter is never
treated as a statistical estimation problem and no statistical
inference is routinely done for unknown reasons, at least in the
frequentist's approach.

Another inherent problem with both BSS and regularization is the
post-selection inference. Conventional statistical inference is
made on the final model with selected variables or nonzero
coefficients by ignoring the effect of model selection, which can
be problematic as pointed out by \citet{Leeb:2005} among others.
One evidence is that no statistical inference is available for
parameters associated with those unselected variables in BSS or
zero estimates in regularization. How to make valid post-selection
inference is currently under intensive statistical research. See,
e.g., \citet{Berk:2013}, \citet{Efron.2014}, and
\citet{Lockhart:2014}.

In this article, we propose a new sparse estimation method for
GLM, termed Minimum approximated Information Criterion (MIC). The
main idea is to reformulate the problem by approximating the
$\ell_0$ norm in (\ref{best-subset}) with a continuous function.
This leads to a smoothed version of BIC that can be directly
optimized. We then devise a reparameterization step that helps
enforce sparsity in parameter estimates while maintaining
smoothness of the objective function at the same time. The
formulation results in a non-convex yet smooth programming
problem. This setup allows us to borrow strength from established
methods and theories in both optimization and statistical
estimation. Many available smooth optimization algorithms can be
conveniently used to solve MIC. At the same time, the smoothness
of the estimating equation allows us to derive valid significance
testings on parameters that are free of post-selection inference.

Our proposed MIC method combines model selection and parameter
estimation together under the common framework of optimization and
accomplishes both within one single step. Compared to many
currently available methods, it offers the three major advantages.
First, MIC yields the best performance to date in sparse
estimation with fixed dimensions because it seeks optimization of
BIC, albeit approximated, without reducing the search space.
Secondly, MIC is computationally advantageous by avoiding
selection of the tuning parameters. Thirdly, MIC makes available
inference results for both zero and non-zero coefficient estimates
via the reparameterization trick. MIC was first proposed by
\citet{Su.2015} in linear regression with focus on variable
selection only.

We emphasize again all our discussions are restricted to fixed
dimensions. The remainder of this article is organized as follows.
Section \ref{sec2} presents the MIC method in detail. In Section
3, we explore its asymptotic properties under regular conditions.
Section 4 presents simulation studies and data analysis examples.
Section 5 ends the article with a brief discussion.

\section{Minimizing the Approximated BIC}
\label{sec2} Our proposed method conducts sparse estimation of GLM
by minimizing an approximated Bayesian information criterion. In
its final form, MIC simply solves the following unconstrained
smooth optimization problem:
\begin{equation}
\min_{\displaystyle \bfgamma} ~~~ - 2 \, L(\mathbf{W} \bfgamma)
~+~ \log(n) \cdot \mbox{tr}(\mathbf{W}), \label{SU00}
\end{equation}
where $\bfbeta =\mathbf{W} \bfgamma$ and $\mathbf{W} =
\mbox{diag}\left(w_j \right)$ with $w_j = w(\gamma_j) = \tanh (a
\, \gamma_j^2 )$ for $j=1, \ldots, p.$ The formulation of
(\ref{SU00}) involves a nonnegative parameters $a$, which controls
the sharpness of approximation. Although asymptotic results
suggest $a = O(n)$, the empirical performance of MIC is rather
stable with respect to the choice of $a$. Thus $a$ will be fixed
{\it a priori}.

The MIC method in (\ref{SU00}) can be described in two steps: (i.)
approximating cardinality with a unit dent function and (ii.)
achieving sparsity with reparameterization. We shall explain the
detailed procedure step-by-step in the ensuing subsections.

\subsection{Unit Dent Functions}
\label{sec-penalties} First of all, we seek an approximation to
the cardinality in (\ref{best-subset}) with a continuous or smooth
surrogate function $w(\cdot)$. This will make the discrete
optimization problem in (\ref{best-subset}) continuous. For the
convenience of presentation, we shall use $\beta$ as a generic
notation for $\beta_j$ from time to time. The cardinality of
$\bfbeta$ is $\sum I \{\beta_j \neq 0 \}$ and hence it reduces to
approximating the indicator function $I\{ \beta \neq 0 \}.$ To
this end, a suitable surrogate function $w(\beta)$ must be a unit
dent function, as defined below.
\begin{definition} \label{def-unit-dent}
Denote $\bar{\mathbb{R}}=R\cup \{-\infty,\infty\}$. A \textit{unit
dent function} is a continuous function $w: \bar{\mathbb{R}}
\rightarrow [0,1]$ that satisfies the following properties:
\begin{enumerate}[(i)]
\item $w(\cdot)$ is an even function such that $w(\beta) =
w(-\beta);$

\item $w(0) =0 $ and $\lim_{\beta \rightarrow \infty} w(\beta)
=1$;

\item $w(\beta)$ is increasing on $\mathbb{R}_+$.
\end{enumerate}
\end{definition}
The above definition implies that $w(\beta)$ is decreasing on
$\mathbb{R}_-$ and $\lim_{\beta \rightarrow -\infty} w(\beta) =1.$
If $w(\beta)$ is differentiable, then $\dot{w}(\beta) \geq 0$ on
$\mathbb{R}+$ and $\dot{w}(\beta) \leq 0$ on $\mathbb{R}_-$. The
$[0,1]$ range requirement essentially makes $w(\cdot)$ non-convex,
but this is necessary in order for $\sum w(\beta)$ to approximate
cardinality, namely, $\parallel \! \bfbeta \! \parallel_0 \approx
\, \sum_{j=1}^p w(\beta_j).$ In addition, the condition
$\lim_{|\beta| \rightarrow \infty} w(\beta) =1$ implies that
$w(\beta)$ is approximately a constant function and hence
$\dot{w}(\beta) =0$ for large $\beta$. As a consequence, when used
as a penalty function, $w(\beta)$ essentially does not alter the
related normal equations or score equations. Motivated by bump
functions, we name $w(\cdot)$ a `dent' function. A special family
of bump functions, called mollifiers, are known as smooth
approximations to the identity \citep{Friedrichs.1944}. If a
mollifier $\phi(\cdot)$ is normalized to have the range $[0,1]$,
then $1-\phi(\cdot)$ is a unit dent function.

\begin{landscape}
\begin{figure}[tp]
\centering
  \includegraphics[scale=0.6, angle=270]{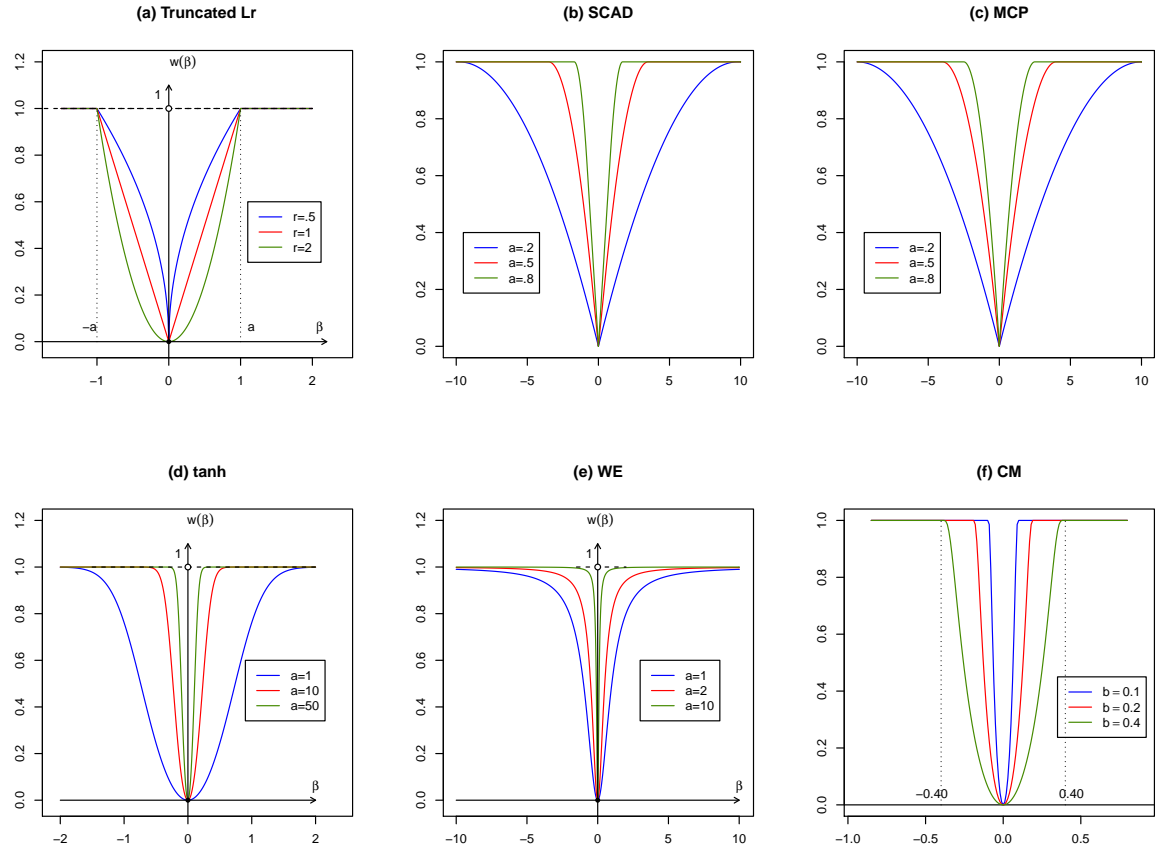}
  \caption{Several unit dent functions for approximating $I(\beta \neq
  0).$ (a) Truncated $L_r$:  $w(\beta; a, r)= \left(|\beta|/a\right)^r$
if $|\beta| \leq a$ and 1 otherwise; (b) modified SCAD: $w(\beta;
a) = a |\beta|$ if $|\beta| \leq a$; $\{2a(2-a^2)|\beta| - a^4 -
a^2 \beta^2 \}/\{4(1-a^2)\}$ if $a < |\beta| < (2-a^2)/a$; and 1
if $|\beta| > (2-a^2)/a$ for $0 < a < \sqrt{2/3}$; (c) modified
MCP: $w(\beta; a) = a|\beta| - a^2 \beta^2/4$ if $|\beta| \leq
2/a$ and 1 if $|\beta| > 2/a$ for $0 < a < \sqrt{2}$; (d)
hyperbolic tangent $w(\beta; a) = \tanh(a \cdot\beta^2);$ (e)
Weight Elimination (WE);  (f) Converse Mollifier (CM) $ w(\beta) =
1 - \exp\left\{- \beta^2/(b^2 - \beta^2)\right\} \cdot
I\left\{|\beta| \leq b \right\}$ for $b
>0$. \label{fig1}}
\end{figure}
\end{landscape}

Let $\mathcal{D}$ denote the family of all unit dent functions. It
can be easily seen that $\mathcal{D}$ is closed under operations
such as composition and product. In particular, it is closed under
power transformation. Namely, if $w(\beta) \in \mathcal{D}$, then
$w^k(\beta) \in \mathcal{D}$ for $k \in \mathbb{N}$. It is worth
noting that unit dent functions have appeared in the
regularization literature. These include the truncated $\ell_r$
penalty studied by \citet{Shen.etal.2012}. The penalty functions
SCAD \citep{Fan.Li.2001} and MCP \citep{Zhang.2010} can also be
modified into unit dent functions. See Figure \ref{fig1} for
graphical illustrations of a number of unit dent functions.

To enforce sparsity, it is necessary for the penalty function to
be non-smooth with a singularity at $\beta=0$, as indicated by
\citet{Fan.Li.2001}. In our proposal, however, we advocate the use
of smooth unit dent functions. The primary reason is that we want
the proposed method to be a natural extension of maximum
likelihood estimation. Since most likelihood or log-likelihood
functions are smooth, we do not want to alter this nature.
Furthermore, the smoothness property allows us to capitalize on
well-developed theories and methods in both optimization and
statistical inference. Our approach is to have smooth penalty
functions and achieve sparsity in a different way.

While many smooth unit dent functions can be considered, we shall
mainly focuses on the hyperbolic tangent function,
\begin{equation}
w(\beta) \, =\, \tanh(a \beta^2) \, =\, \frac{\exp(2 a
\beta^2)-1}{\exp(\displaystyle 2 a \beta^2)+1} \, =\, 2 \,
\mbox{logistic}(2a \beta^2) -1. \label{tangent2}
\end{equation}

\noindent This is because its derivatives are easily calculated,
with the first two given by $\dot{w}(\beta) ~ = ~ 2 a \beta
(1-w^2)$ and $\ddot{w}(\beta) ~ = ~ 2a(1-w^2)(1-4a\beta^2 w).$ In
addition, the $\tanh(\cdot)$ function is associated with the
logistic or expit function which is widely used in statistics. A
plot of $w(\beta)$ versus $\beta$ for different $a$ values is
provided in Figure \ref{fig1}(d). It can be seen that a larger $a$
yields a sharper approximation to the indicator function $I\{\beta
\neq 0 \}.$

\begin{figure}[h]
\centering
  \includegraphics[scale=0.55, angle=270]{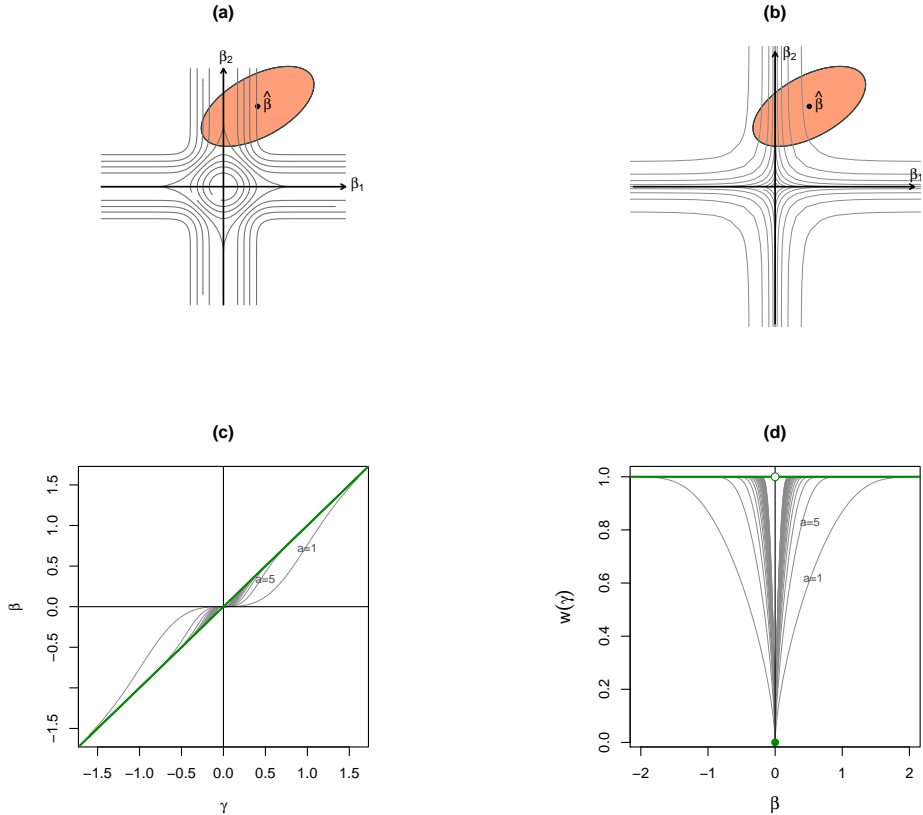}
  \caption{Illustration of the reparameterization step:
  (a) the contour plot for the optimization problem in
  the two-dimensional scenario before reparameterization; (b)
  the contour plot after reparameterization; (c) $\beta = \gamma w(\gamma)$ vs.~$\gamma$;
  and (d) $w(\gamma)$ as a penalty function for $\beta$. Different $a$ values in \{1, 5, 10, 15,
  \ldots, 100\} are used .
 \label{fig02}}
\end{figure}

With the surrogate function $w(\beta)=\tanh(a \beta^2)$, we seek
to solve
\begin{equation}
\min_{\bfbeta}~~ - 2 \cdot L(\bfbeta) + \lambda_0 \cdot
\sum_{j=1}^p w(\beta_j). \label{SU-I}
\end{equation}
Expanding $L(\bfbeta)$ at the MLE $\widehat{\bfbeta}$ and then
using the fact  that $ \nabla L (\widehat{\bfbeta}) = \mathbf{0}$,
we have
$$ L(\bfbeta) \, \approx \, L(\widehat{\bfbeta}) + (\bfbeta -
\widehat{\bfbeta})^T \left\{\nabla^2 L (\widehat{\bfbeta}) /2
\right\} (\bfbeta - \widehat{\bfbeta}),$$ where $\nabla L
(\widehat{\bfbeta})$ and $\nabla^2 L (\widehat{\bfbeta})$ are the
gradient vector and Hessian matrix of $L(\bfbeta)$ evaluated at
$\widehat{\bfbeta}$, respectively. Thus, the penalized
optimization form in (\ref{SU-I}) can be viewed as the Lagrangian
that roughly corresponds to a constrained optimization problem:
\begin{equation} \min_{\bfbeta}~~ (\bfbeta - \widehat{\bfbeta})^T \left\{ - \, \nabla^2
L (\widehat{\bfbeta}) \right\} (\bfbeta - \widehat{\bfbeta})
\mbox{~~subject to~~} \sum_{j=1}^p w(\beta_j) \leq t_0,
\label{contour}
\end{equation}
for some $t_0 \geq 0$. Figure \ref{fig02}(a) presents a graphical
illustration of the optimization problem (\ref{contour}) in the
two-dimensional case. The objective function in (\ref{contour}) is
an ellipsoid centered at MLE $\widehat{\bfbeta}$. The feasible set
for the constraint $w(\beta_1) + w(\beta_2) \leq t_0$ contains
both sharpened diamonds for large $t_0$ and discs for small $t_0$
as shown in the contour plots of Figure \ref{fig02}(a). By the
Taylor expansion, $w(\beta) = a \beta^2 + O(\beta^6)$ for $\beta
\rightarrow 0$. Thus, it is not surprising that $w(\beta)$ behaves
similarly to the ridge $\ell_2$ penalty around 0. This implies
that sparsity may not be enforced. We shall address this issue in
the next section. Hereafter, we consider $w(\beta)$ to be the
hyperbolic tangent penalty, unless otherwise explicitly stated.

\subsection{Reparameterization}
\label{section-uprooting}

To enforce sparsity, we consider a reparameterization procedure
originally motivated from the nonnegative garrotte (NG) of
\citet{Breiman.1995}. NG can be viewed as a sign-constrained
regularization that is based on the decomposition $\beta =
\sgn(\beta) |\beta|$. Supposing that the sign of each $\beta_j$
can be correctly specified by the MLE $\widehat{\bfbeta}$, it
remains to estimate $|\bfbeta_j|$. Reparameterizing $\bfbeta =
\diag(\widehat{\bfbeta}) \, \bfgamma$ for some nonnegative vector
$\bfgamma$ such that $\gamma_j =|\beta_j|$ leads to the NG
formulation
$$ \min_{\bfgamma} - 2 L(\bfbeta) \mbox{~~s.t.~} \sum_{j=1}^p
\gamma_j \leq t \mbox{~and~} \gamma_j \geq 0,$$ where $t$ is a
tuning parameter. One fundamental problem with sign-constrained
regularization is that if any sign is wrongly specified by the
initial estimator $\widehat{\bfbeta}$, which occurs often with
real data owing to multicollinearity or other complexities, then
it is not possible to make correction.

Our immediate aim is to introduce singularity to the penalty
function at 0. For this purpose, we consider the decomposition
$\beta = \beta \, I\{\beta \neq 0 \}$. Set $\gamma = \beta$ and
approximate $I\{\gamma \neq 0 \}$ with $w(\gamma)$. Namely,
$$ \gamma \, = \, \gamma \, I\{\gamma \neq 0 \} ~ \approx ~ \gamma
w(\gamma).$$ This motivates the reparameterization $\beta_j$ as
$\beta_j = \gamma_j w(\gamma_j)$ for $j=1, \ldots, p.$ In matrix
form, $\bfbeta = \mathbf{W} \bfgamma,$ where matrix $\mathbf{W}$
is defined earlier in (\ref{SU00}). As shown in Figure
\ref{fig02}(c), $\beta$ is an strictly increasing function of
$\gamma$ and $\beta = \gamma$ except for a small neighborhood of
0, in which a shrinkage on $|\beta|$ is imposed.

To see how the reparameterization helps with enforcing sparsity,
consider the resulting optimization problem:
\begin{equation}
\min_{\bfbeta}~~ - 2 \cdot L(\bfbeta) + \ln(n) \cdot \sum_{j=1}^p
w(\gamma_j). \label{SU-II}
\end{equation}
Compared to (\ref{SU-I}), the only change is that the penalty
function $w(\cdot)$ is now applied to the reparameterized
$\gamma_j$ instead of $\beta_j.$ It is worth noting that the
penalty function $w(\gamma_j)$ in (\ref{SU-II}) is an implicit
function of $\beta_j$. Figure \ref{fig02}(d) plots $w(\gamma)$ as
a penalty function of $\beta$ for different values of $a,$ which
now shows a similar pattern to the non-convex SCAD or MCP penalty
with a cusp at $\beta=0$. It can be easily verified that
$w(\gamma)$ remains a unit dent function of $\beta$ that well
approximates $I(\beta \neq 0).$

The singularity at 0 can be further confirmed by calculating the
derivatives of $w(\gamma)$ at $\beta$. Applying the chain rule
gives
\begin{equation}
\frac{d\, w(\gamma)}{d \, \beta} = \frac{d\, w(\gamma)}{d \,
\gamma} \cdot \frac{d\, \gamma}{d \, \beta} = \frac{d\,
w(\gamma)}{d \, \gamma} \cdot \left( \frac{d\, \beta}{d \, \gamma}
\right)^{-1} = \frac{\displaystyle \dot{w}}{\displaystyle w +
\gamma \dot{w}} , \label{first-derivative}
\end{equation}
where we denote $w = w(\gamma)$ and $\dot{w} = \dot{w}(\gamma) = 2
a \gamma (1-w^2),$ and it follows $d\, \beta / d \, \gamma = w +
\gamma \dot{w}.$ The first derivative in (\ref{first-derivative})
is expressed in terms of $\gamma$ via implicit differentiation
since the explicit formula of $\gamma$ in terms of $\beta$ is
unavailable. The validity of (\ref{first-derivative}), however,
requires $d\, \beta / d \, \gamma ~\neq ~0$, which holds
everywhere except at $\beta=0.$ Similar arguments can be used to
derive the form of the higher-order derivatives. For example, the
second-derivative is given by $$\frac{\displaystyle d^2\,
w(\gamma)}{\displaystyle d\, {\beta}^2} = \frac{\displaystyle w \,
\ddot{w} - 2 \, \dot{w}^2} { \displaystyle (w + \gamma \dot{w} )^3
}$$ with $\ddot{w} = \ddot{w}(\gamma) = 2 a(1-w^2)(1-4a \gamma^2
w),$ which again does not exist at $\beta=0.$ It can be verified
that $w(\gamma)$ is a smooth function of $\beta$ except at
$\beta=0.$

It is worth mentioning that the property that the
reparameterization $\beta = \gamma w(\gamma)$ helps enforce
singularity at 0 holds for any smooth function in $\mathcal{D}$.
We have utilized the differentiation of the inverse function to
achieve this. Accordingly, the derivatives of $w(\gamma)$ as a
function of $\beta$ exist everywhere except when $\beta = 0$.
There should be other ways of introducing singularities for smooth
functions.

Figure \ref{fig02}(b) provides a two-dimensional illustration of
the constrained optimization version that corresponds to
(\ref{SU-II}):
\begin{equation*}
\min_{\bfbeta}~~ (\bfbeta - \widehat{\bfbeta})^T \left\{ - \,
\nabla^2 L (\widehat{\bfbeta}) \right\} (\bfbeta -
\widehat{\bfbeta}) \mbox{~~subject to~~} \tr(\mathbf{W}) \leq t_0
\mbox{~with~} \bfbeta = \mathbf{W} \bfgamma.
\end{equation*}
The contour lines of the constraint $w(\gamma_1) + w(\gamma_2)
\leq t$ (as a function of $\beta_1$ and $\beta_2$ now) become
sharpened diamonds, which serves better for the variable selection
purpose.

Besides achieving sparsity, the smooth formulation facilitated by
reparameterization allows us to further capitalize on available
results in both optimization theory and statistical inference and
leads to some important conveniences and advantages. For the
computation purpose, we shall estimate $\gamma$ instead by solving
(\ref{SU00}). Compared to (\ref{SU-II}) where the objective
function is nonsmooth in $\bfbeta$, we have now switched the
decision vector to $\bfgamma$ instead of $\bfbeta.$ Solving
(\ref{SU00}) is a smooth optimization problem and many standard
algorithms can be applied. Estimation of $\bfgamma$ is meaningful
in its own right. The fact that the correspondence between $\beta$
and $\gamma$ is one-to-one with $\beta_j=0$ iff $\gamma_j = 0$
allows us to derive significance testing results for $\bfbeta$
through $\bfgamma$, which are free of post-selection inference.
The objective function in (\ref{SU00}) is smooth for estimating
$\bfgamma.$ Thus standard arguments in M-estimators can be applied
for obtaining inference on $\bfgamma.$ The detailed procedure will
be explained in next section.

\section{Asymptotic Properties}
\label{sec-asymptotics}

In this section, we first study the asymptotic oracle properties
of the MIC estimator $\widetilde{\bfbeta}$, including its
$\sqrt{n}$-consistency, selection consistency, and the asymptotic
normality of its nonzero components. We then present significance
testing on $\bfbeta$ via $\bfgamma$, which is free of
post-selection inference. Once again, we emphasize that all our
discusses are restricted to the fixed $p$ scenarios.

\subsection{Oracle Properties of the MIC Estimator $\widetilde{\bfbeta}$}

For theoretical investigation, we consider the MIC estimator
$\widetilde{\bfbeta}$ obtained from minimizing the objective
function in (\ref{SU-II})
\begin{equation}
Q_n(\bfbeta) =  - 2 \, \frac{L(\bfbeta)}{n} + \frac{\ln(n)}{n} \,
\sum_{j=1}^p w(\gamma_j), \label{Qn}
\end{equation}
where $ L(\bfbeta)=\sum_{i=1}^n l_i(\bfbeta)$ with $l_i(\bfbeta) =
\log f(\X_i, Y_i; \bfbeta)$. We shall denote $a$ as $a_n$ so that
$\beta_j= \gamma_j w(\gamma_j) = \gamma_j \tanh(a_n \beta_j^2)$
and assume $a_n = O(n)$; this rate for $a_n$ will be manifested in
the derivation.

Denote the true parameter as $\bfbeta_0 = ({\bfbeta}_{0(1)}^T,
{\bfbeta}_{0(0)}^T)^T,$ where ${\bfbeta}_{0(1)} \in
\mathbb{R}^{q}$ consists of all $q$ nonzero components and
${\bfbeta}_{0(0)} = \mathbf{0}$ consists of all the $(p-q)$ zero
components. As generic notation, we use $\widetilde{\bfbeta}$ and
$\widehat{\bfbeta}$ to denote the MIC and MLE estimators,
respectively. Let $\mathbf{I}=\mathbf{I}(\bfbeta_0)$ be the
expected Fisher information matrix for the whole model and let
$\mathbf{I}_1$ be the Fisher information corresponding to the
reduced true model setting $\bfbeta_{0(0)} =\mathbf{0}.$ It is
well known that $\mathbf{I}_1$ is the $q$-th principal submatrix
of $\mathbf{I}.$ The following theorem shows that, under
regularity conditions, there exists a local minimizer
$\widetilde{\bfbeta}$ of $Q_n(\bfbeta)$ that is
$\sqrt{n}$-consistent and this $\sqrt{n}$-consistent
$\widetilde{\bfbeta}$ enjoys the `oracle' property.

\begin{thm}
Let $\{(\X_i, Y_i): i=1, \ldots, n\}$ be $n$ i.i.d. copies from a
density $f(\X, Y; \bfbeta_0).$ Under the regularity conditions
(A)--(C) in \citet{Fan.Li.2001}, we have
\begin{enumerate}[(i)]
\item[(i).] ($\sqrt{n}$-Consistency)~~ there exists a local
minimizer $\widetilde{\bfbeta}$ of $Q_n(\bfbeta)$ that is
$\sqrt{n}$-consistent for $\bfbeta_0$ in the sense that $\parallel
\widetilde{\bfbeta} - \bfbeta_0 \parallel = O_p (n ^{-1/2}).$

\item[(ii).] (Sparsity and Asymptotic Normality)~~ Partition
$\widetilde{\bfbeta}$ in (i) as  $(\widetilde{\bfbeta}_{(1)}^T,
\widetilde{\bfbeta}_{(0)}^T)^T$ in a similar manner to
$\bfbeta_0$. With probability tending to 1 as $n \rightarrow
\infty$, $\widetilde{\bfbeta}$ must satisfy that
$$\widetilde{\bfbeta}_{(0)} = \mathbf{0}$$ and
$$ \sqrt{n}(\widetilde{\bfbeta}_{(1)} - \bfbeta_{0(1)}) \, \rightarrow
\, N\left(\mathbf{0}, \, \mathbf{I}_{1}^{-1} \right).$$
\end{enumerate}
\label{theorem-SCAD}
\end{thm}

The results in Theorem \ref{theorem-SCAD} are analogous to
Theorems 1 \& 2 in \citet{Fan.Li.2001}. It establishes that
$\widetilde{\bfbeta}_{(0)}$ is selection consistent and
$\widetilde{\bfbeta}_{(1)}$ is a best asymptotic normal (BAN; see,
e.g., \citeauthor{Serfling.1980}, \citeyear{Serfling.1980})
estimator of $\bfbeta_{0(1)}.$ We defer its proof to the Appendix.
The standard errors (SE) for nonzero components in
$\widetilde{\bfbeta}$ can be conveniently computed by replacing
$\mathbf{I}_1$ in Theorem \ref{theorem-SCAD}(ii) with the observed
Fisher information matrix \citep{Efron.1978} and plugging in
$\widetilde{\bfbeta}$. Since $\widetilde{\bfbeta}$ is essentially
an M-estimator, alternative sandwich SE formulas
\citep{Stefanski.2002} are available, for which we shall not
pursue further. However, as post-selection inferences, all these
SE formulas are only available for nonzero components in
$\widetilde{\bfbeta}$ and hence caution should be exercised.


\renewcommand{\tabcolsep}{5pt}
\renewcommand{\arraystretch}{1.2}
\renewcommand{\baselinestretch}{1}
\begin{table}
\begin{small}
\begin{center}
\caption{Simulation results on MIC (with $\lambda_0=\ln(n)$ and
$a=10$) in comparison with other methods. Reported quantities
include the averaged model errors (ME), the averaged model size
(Size), the average number of false positive variables (FP), the
average number of false negative variables (FN), the proportion of
correct selections (C), all based on 500 realizations.}
\label{tbl-simu} \centering \vspace{.1in}
\begin{tabular}{lccccccccccc}
\multicolumn{12}{c}{(a) Model A -- Linear Regression} \\ \hline
       &   \multicolumn{5}{c}{$n=100$}                                   &&  \multicolumn{5}{c}{$n=200$}     \\   \cline{2-6} \cline{8-12}
Method  &        ME  &   Size    &   FP  &   FN  &   C    &&  ME
&   Size    &   FP  &   FN  &   C   \\  \hline
MIC   &   0.054   &   3.47    &   0.47    &   0.00    &   0.640   &&  0.021   &   3.25    &   0.25    &   0.00    &   0.790   \\
Oracle      &   0.034   & 3.00      & 0.00      & 0.00      & 1.000     &&  0.015   & 3.00      & 0.00      & 0.00      & 1.000     \\
BIC   &        0.055   &   3.35    &   0.35    &   0.00    &   0.710   &&  0.022   &   3.19    &   0.19    &   0.00    &   0.834   \\
LASSO         &   0.085   &   6.09    &   3.09    &   0.00    &   0.092   &&  0.039   &   6.23    &   3.23    &   0.00    &   0.102   \\
SCAD          &   0.045   &   3.58    &   0.58    &   0.00    &   0.752   &&  0.022   &   3.71    &   0.71    &   0.00    &   0.752   \\
MCP       &   0.047   &   3.57    &   0.57    &   0.00    & 0.750 &&  0.020   &   3.41    &   0.41    &   0.00    &   0.814 \\
\hline
\end{tabular}

\vspace{.2in}
\begin{tabular}{lccccccccccc}
\multicolumn{12}{c}{(b) Model B -- Logistic Regression} \\ \hline
       &   \multicolumn{5}{c}{$n=100$}                                   &&  \multicolumn{5}{c}{$n=200$}                                   \\   \cline{2-6} \cline{8-12}
Method  &         ME  &   Size    &   FP  &   FN  &   C    && ME
&   Size    &   FP  &   FN  &   C   \\  \hline
MIC    &   0.017   &   3.74    &   1.03    &   0.29    &   0.354   &&  0.005   &   3.42    &   0.49    &   0.07    &   0.624   \\
Oracle      &   0.005   & 3.00      & 0.00      & 0.00      & 1.000     &&  0.002   & 3.00      & 0.00      & 0.00      & 1.000     \\
BIC  &         0.015   &   3.40    &   0.67    &   0.27    &   0.514   &&  0.005   &   3.21    &   0.28    &   0.06    &   0.766   \\
LASSO   &         0.023   &   6.54    &   3.79    &   0.25    &   0.012   &&  0.012   &   7.32    &   4.37    &   0.05    &   0.018   \\
SCAD    &         0.019   &   3.69    &   1.09    &   0.41    &   0.206   &&  0.012   &   3.92    &   1.11    &   0.19    &   0.278   \\
MCP &         0.019   &   3.12    &   0.65    &   0.53    & 0.236 &&  0.011   &   3.39    &   0.64    &   0.24    &   0.420 \\
\hline
\end{tabular}

\vspace{.2in}
\begin{tabular}{lccccccccccc}
\multicolumn{12}{c}{(c) Model C -- Log-Linear Regression} \\
\hline
      &   \multicolumn{5}{c}{$n=100$}                                   &&  \multicolumn{5}{c}{$n=200$}                                   \\  \cline{2-6} \cline{8-12}
Method       &   ME  &   Size    &   FP  &   FN  &   C    &&  ME
&   Size    &   FP  &   FN  &   C   \\  \hline
MIC  &   12.310  &   3.34    &   0.35    &   0.00    &   0.712   &&  4.367   &   3.23    &   0.23    &   0.00    &   0.828   \\
Oracle      &   9.289   & 3.00      & 0.00      & 0.00      & 1.000     &&  3.555   & 3.00      & 0.00      & 0.00      & 1.000     \\
BIC  &          25.884  &   3.39    &   0.39    &   0.00    &   0.714   &&  4.897   &   3.23    &   0.23    &   0.00    &   0.826   \\
LASSO   &       600.821 &   1.55    &   0.37    &   1.81    &   0.184   &&  348.182 &   1.46    &   0.18    &   1.72    &   0.282   \\
SCAD    &       40.753  &   4.08    &   1.08    &   0.00    &   0.336   &&  12.843  &   3.64    &   0.64    &   0.00    &   0.528 \\
\hline
\end{tabular}
\end{center}
\end{small}
\end{table}

\subsection{Inference on $\bfbeta$ via $\bfgamma$}
\label{sec-inference}

MIC avoids the two-step estimation process in the best subset
selection and regularization by completing both variable selection
and parameter estimation in one single optimization step. This
brings about a unique opportunity to address the fundamental
post-selection inference problem.

Inference on zero components in $\bfbeta$ is unavailable in MIC.
This is because asymptotic normality of M-estimators often
involves a condition that the expected objective function
$E\{Q_n(\bfbeta)\}$ admits a second-order Taylor expansion at
$\bfbeta_0$ whereas sparsity requires singularity of the penalty
function $w(\gamma)$ as a function of $\beta$ at $\beta=0.$
However, the reparameterisation helps us to circumvent this
non-smoothness issue. The transformation $\beta= \gamma w(\gamma)$
is a bijection and $\beta=0$ iff $\gamma=0.$ Therefore, testing
$H_0:~\beta_j=0$ is equivalent to testing $H_0: \gamma_j =0.$ As
the objective function of $\bfgamma$, $Q_n(\bfgamma)$ in
(\ref{Qn}) is smooth in $\bfgamma.$ Therefore, the statistical
properties of $\widetilde{\bfgamma}$ are readily available
following standard M-estimation arguments, as given in the theorem
below.
\begin{thm}
Let $\bfgamma_{0}$ be the reparameterized parameter vector
associated with $\bfbeta_0$ such that $\beta_{0j} = \gamma_{0j}
w(\gamma_{0j}).$ It follows that $\parallel \bfgamma_0 - \bfbeta_0
\parallel_2 = O\{ \exp(-2a_n  \min_{1 \leq j \leq q}
\gamma^2_{0j})\}.$ Under the regularity conditions (A)--(C) in
\citet{Fan.Li.2001}, we have
\begin{equation}
\sqrt{n} \left[ \mathbf{D}(\bfgamma_0) (\widetilde{\bfgamma} -
\bfgamma_0) + \mathbf{b}_n \right] ~ \stackrel{d}{\longrightarrow}
~  N\left\{\mathbf{0}, \, \mathbf{I}^{-1}(\bfbeta_0) \right\}.
\label{asyp-norm-bfgamma}
\end{equation}
where
\begin{equation}
 \mathbf{D}(\bfgamma_0) = \left. \diag(w_j + \gamma_j \dot{w}_j) \right|_{\bfgamma = \bfgamma_0} = \diag \left(D_{jj} \right)
\label{D-gamma0}
\end{equation}
and the asymptotic bias
\begin{equation}
\mathbf{b}_n = \left\{ - \nabla^2 L(\bfbeta_0) \right\}^{-1} \,
\frac{\ln(n)}{2 } \left( \frac{\dot{w}_j}{w_j + \tilde{\gamma}_j
\dot{w}_j} \right)_{j=1}^p \, = \, \left( b_{nj} \right)_{j=1}^p
\label{bias}
\end{equation}
satisfy (i) $\lim_{n \rightarrow \infty} D_{jj}  = I\{ \beta_{0j}
\neq 0 \}$ and (ii) $\mathbf{b}_n = o_p(1).$
\label{theorem-bfgamma}
\end{thm}
The proof of Theorem \ref{theorem-bfgamma} is given in the
Appendix. One practical implication of Theorem
\ref{theorem-bfgamma} is that both $\mathbf{D}(\bfgamma_0)$ and
$\mathbf{b}_n$ may be ignored in computing the standard errors of
$\widetilde{\bfgamma}.$ Furthermore, since $ \parallel
\widetilde{\bfgamma} - \bfbeta_0
\parallel \, \leq \,
\parallel \widetilde{\bfgamma} - \bfgamma_0 \parallel + \parallel
\bfgamma_0 - \bfbeta_0 \parallel = o_p(1),$ $\widetilde{\bfgamma}$
is a consistent estimator of $\bfbeta_0$ and can be used to
replace $\bfbeta_0$ in estimating the Fisher information matrix.
Thus, an asymptotic $(1-\alpha) \times 100\%$ confidence interval
for $\gamma_{0j}$ can be simply given by
\begin{equation}
\tilde{\gamma}_j \, \pm \, z_{1-\alpha/2} \, \sqrt{\left(
\mathbf{I}_n^{-1}(\widetilde{\bfgamma})/n \right)_{jj}},
\label{CI-bfgamma}
\end{equation}
where $\mathbf{I}_n$ denotes the observed Fisher information
matrix and $z_{1-\alpha/2}$ is the $(1-\alpha/2)$-th percentile of
$N(0,1).$ Significance testing on $\gamma_{0j}$ can be done
accordingly. There are alternative ways to derive the asymptotic
variance of $\widetilde{\bfbeta}$. We numerically experimented a
couple of other sandwich estimators and found that the simple
formula in (\ref{CI-bfgamma}) performs very well empirically.

\section{Numerical Results}

In this section, we present simulation experiments and real data
examples to illustrate MIC in comparison with other methods.

\subsection{Computational Issues}

MIC solves for $\widetilde{\bfgamma}$ by optimizing (\ref{SU00}).
Considering its nonconvex nature, a global optimization method is
desirable. \citet{Mullen.2014} provides a comprehensive comparison
of many global optimization algorithms currently available in R
\citep{R.2016}. According to her recommendations, we have chosen
the \texttt{GenSA} package \citep{Xiang.2013} that implements the
generalized simulation annealing of \citet{Tsallis.1996}, because
of its superior performance in both identification of the true
optimal point and computing speed. With estimated
$\widetilde{\bfgamma}$, the MIC estimator $\widetilde{\bfbeta}$ of
$\bfbeta$ can be obtained immediately via the transformation
$\widetilde{\bfbeta} = \widetilde{\mathbf{W}}
\widetilde{\bfgamma}$, where $\widetilde{\mathbf{W}}= \diag
(\tilde{w}_j)$ with $\tilde{w}_j = w(\tilde{\gamma}_j).$ Because
of the shrinkage effect of the reparameterization around 0,
estimates $\tilde{\gamma}_j$ that are close to 0 would yield very
small values of $\tilde{\beta}_j$, which can be virtually taken as
0.

Implementation of MIC involves of the choice of $a_n$. In theory,
the asymptotic results in Section \ref{sec-asymptotics} entail
that $a_n = O(n).$ In order to apply the arguments of
\citet{Fan.Li.2001}, this $O(n)$ rate seems unique. See the proofs
of Theorem \ref{theorem-SCAD} in the appendix. On the other hand,
if one is willing to adjust the choice of $\lambda_0$, recall that
the selection consistency of BIC holds for a wide range of
$\lambda$ values, then the choice of $a_n$ can be more flexible.
However, the conventional choice of $\lambda_0 = \ln(n)$ is
optimal in the Bayesian sense \citep{Schwarz.1978}. Thus it is
advisable to keep it as is. In practice, the empirical performance
of MIC stays rather stable with respect to the choice of $a_n,$ as
demonstrated in \citet{Su.2015} for linear regression. The role of
$a$ is quite different from the tuning parameter $\lambda$ in
regularization that controls the penalty for complexity or the
range of certain constraints. When $\lambda$ varies, the parameter
estimates would change dramatically, which necessitates selection
of $\lambda.$ In MIC, $a_n$ is a shape or scale parameter in the
unit dent function that modifies the sharpness of its
approximation to the indicator function. The role of $a_n$ is
largely similar to that of the parameter $a$ in SCAD
\citep{Fan.Li.2001}, where $a$ is fixed as $a=3.7.$ In general, a
larger $a_n$ value enforces a better approximation of the
indicator function with the hyperbolic tangent function. On the
other hand, a smaller $a_n$ is appealing for optimization
purposes, by introducing more smoothness. Based on our numerical
experiences, applying a $a$ value smaller than 1 leads to less
stable and reduced performance. The performance of MIC stabilizes
substantially when $a_n$ gets large, especially when it is 10 or
above. On this basis, we recommend choosing $a_n$ at a value in
[10, 50] for standardized data. Avoiding tuning $a_n$ makes MIC
computationally advantageous.

Four known methods are included for comparison with MIC: the best
subset selection (BSS) with BIC, LASSO, SCAD, and MCP. The oracle
estimate is also added as a benchmark. All the computations are
done in R (R Development Core Team, 2015). Specifically, we have
used the R package \texttt{bestglm} for BSS, \texttt{lars} and
\texttt{glmnet} for LASSO, and \texttt{ncvreg} and \texttt{SIS}
for SCAD and MCP. The default settings are used in these
implementations, presuming that the default setting is the most
recommendable.

\renewcommand{\tabcolsep}{4pt}
\renewcommand{\arraystretch}{1}
\renewcommand{\baselinestretch}{1}
\begin{table*}
\begin{center}
\begin{small}
\caption{Simulation results on standard errors of nonzero
$\widehat{\bfbeta}$ with $n=200$ over 500 simulation runs.
Reported quantities are MAD of the parameter estimates, Median of
the standard errors, and MAD of the standard errors.}
\label{tbl-se} \vspace{.1in} \centering
\begin{tabular}{cccccccccccc}
\multicolumn{12}{c}{(a) Model A -- Gaussian Linear Regression} \\ \hline
  &  \multicolumn{3}{c}{oracle} &&     \multicolumn{3}{c}{MIC}       &&  \multicolumn{3}{c}{Best Subset}   \\  \cline{2-4}  \cline{6-8}  \cline{10-12}
Parameter  &    MAD & Median SE  & MAD SE  &&    MAD & Median SE
& MAD SE   &&   MAD & Median SE &  MAD SE  \\ \hline
$\beta_1$ & 0.083 &  0.082 &   0.006  &&       0.083 &   0.082 &   0.006  &&       0.084 &   0.082 &   0.004   \\
$\beta_2$ & 0.084 &  0.082 &   0.006  &&       0.087 &   0.082 &   0.006  &&       0.085 &   0.082 &   0.004   \\
$\beta_5$ & 0.072 &  0.072 &   0.005  &&       0.073 &   0.072 &   0.005  &&       0.075 &   0.072 &   0.004   \\ \hline
\end{tabular}

\vspace{.1in} \centering
\begin{tabular}{cccccccccccc}
\multicolumn{12}{c}{(b) Model B -- Logistic Regression} \\ \hline
 &   \multicolumn{3}{c}{oracle}&&              \multicolumn{3}{c}{MIC}       &&  \multicolumn{3}{c}{Best Subset}   \\  \cline{2-4}  \cline{6-8}  \cline{10-12}
Parameter  &    MAD & Median SE  & MAD SE  &&    MAD & Median SE
& MAD SE   &&   MAD & Median SE &  MAD SE  \\ \hline
$\beta_1$ & 0.528 &   0.475 &   0.086  &&       0.529 &   0.492 &   0.094  &&       0.545 &   0.488 &   0.090   \\
$\beta_2$ & 0.399 &   0.389 &   0.048  &&       0.448 &   0.407 &   0.064  &&       0.435 &   0.400 &   0.057   \\
$\beta_5$ & 0.380 &   0.356 &   0.059  &&       0.405 &   0.367 &   0.061  &&       0.402 &   0.362 &   0.061   \\ \hline
\end{tabular}

\vspace{.1in} \centering
\begin{tabular}{cccccccccccc}
\multicolumn{12}{c}{(c) Model C -- Loglinear Regression} \\ \hline
 &   \multicolumn{3}{c}{oracle}&&              \multicolumn{3}{c}{MIC}       &&  \multicolumn{3}{c}{Best Subset}   \\  \cline{2-4}  \cline{6-8}  \cline{10-12}
Parameter  &    MAD & Median SE  & MAD SE  &&    MAD & Median SE
& MAD SE   &&   MAD & Median SE &  MAD SE  \\ \hline
$\beta_1$ & 0.037 &   0.036 &   0.007  &&       0.037 &   0.036 &   0.007  &&       0.038 &   0.036 &   0.007   \\
$\beta_2$ & 0.039 &   0.039 &   0.007  &&       0.040 &   0.039 &   0.007  &&       0.041 &   0.039 &   0.007   \\
$\beta_5$ & 0.032 &   0.032 &   0.006  &&       0.033 &   0.033 &   0.006  &&       0.033 &   0.033 &   0.006   \\ \hline
\end{tabular}
\end{small}
\end{center}
\end{table*}

\renewcommand{\tabcolsep}{4pt}
\renewcommand{\arraystretch}{1}
\renewcommand{\baselinestretch}{1}
\begin{table*}
\begin{center}
\caption{Hypothesis testing on $\bfgamma_0$ in MIC. Empirical size
and empirical power are obtained at the significance level
$\alpha=0.05$ based on 1,000 simulation runs.} \label{tbl-gamma}
\vspace{.1in} \centering
\begin{tabular}{lcccccccccccccc} \hline \hline
    &             &   \multicolumn{9}{c}{Empirical Size}                                                                  &&  \multicolumn{3}{c}{Empirical Power}                 \\  \cline{3-11} \cline{13-15}
Model    &   $n$   &   $\gamma_3$   &   $\gamma_4$   &   $\gamma_6$   &   $\gamma_7$   &   $\gamma_8$   &   $\gamma_9$   &   $\gamma_{10}$    &   $\gamma_{11}$    &   $\gamma_{12}$    &&  $\gamma_1$   &   $\gamma_2$   &   $\gamma_5$   \\  \hline
A    &   100 &     0.054   &   0.059   &   0.061   &   0.060   &   0.055   &   0.056   &   0.054   &   0.051   &   0.044   &&  1.000   &   1.000   &   1.000   \\
    &   200 &      0.049   &   0.040   &   0.047   &   0.040   &   0.034   &   0.024   &   0.039   &   0.036   &   0.031   &&  1.000   &   1.000   &   1.000   \\ \hline
B   &   100 &      0.054   &   0.059   &   0.061   &   0.060   &   0.055   &   0.056   &   0.054   &   0.051   &   0.044   &&  1.000   &   1.000   &   1.000   \\
    &   200 &      0.049   &   0.040   &   0.047   &   0.040   &   0.034   &   0.024   &   0.039   &   0.036   &   0.031   &&  1.000   &   1.000   &   1.000   \\ \hline
C   &   100 &      0.042   &   0.048   &   0.047   &   0.034   &   0.030   &   0.034   &   0.036   &   0.044   &   0.041   &&  1.000   &   1.000   &   1.000   \\
    &   200 &     0.022   &   0.025   &   0.025   &   0.042   &   0.024   &   0.029   &   0.023   &   0.021   &   0.024   &&  1.000   &   1.000   &   1.000   \\ \hline
\end{tabular}
\end{center}
\end{table*}

\subsection{Simulated Experiments}
We generate data sets from the following three GLM models by using
the same simulation settings as those of Zou and Li (2008).
Specifically, the following three models are used:
\begin{equation}\left\{
\begin{array}{ll}
\mbox{Model A:~} & y|\x ~ \sim ~ N\{\mu(\x), 1\} \mbox{~~with~}  \mu(\x) = \x^T\bfbeta, \\
\mbox{Model B:~} & y| \x  ~ \sim ~ \mbox{Bernoulli}\{\mu(\x)\}   \mbox{~~with~}  \mu(\x) = \mbox{expit}( \x^T \bfbeta), \\
\mbox{Model C:~} & y|\x ~ \sim ~ \mbox{Poisson} \left\{ \mu(\x)
\right\}  \mbox{~~with~}  \mu(\x) = \exp(\x^T\bfbeta),
\end{array} \right.  \label{models}
\end{equation}
where $\bfbeta = (3, 1.5, 0, 0, 2, 0, 0, 0, 0, 0, 0, 0)^T$ in
Models A and  B, and $(1.2, .6, 0, 0, .8, 0, 0, 0, 0, 0, 0, 0)^T$
in Model C. Each data set involves $p=12$ predictors that follow a
multivariate normal distribution $N(\mathbf{0}, \, \bfSigma)$ with
$\bfSigma=(\sigma_{j j'})$ and $\sigma_{j j'}=0.5^{|j-j'|}$ for
$j, j'=1, \ldots, p.$ In Model B, six binary predictors are
created by setting $x_{2j-1}:= I(x_{2j-1} <0)$ for $j=1, \ldots,
6.$ Thus, there are six continuous and six binary predictors in
Model B. Each simulation includes two different sample sizes
$n=100$ and $n=200$, and 500 realizations are generated from each
model.

To apply the MIC method, we fix $\lambda_0= \ln(n)$ and $a_n=10.$
Five performance measures are used for making comparisons. The
first one is the empirical model error (ME), defined as $
\mbox{ME} = \sum_{i=1}^n (\mu_i - \hat{\mu}_i)^2/n$, where $\mu_i$
is given in (\ref{models}) and $\hat{\mu}_i$ is obtained by
plugging in the estimate of $\bfbeta$. We compute ME based on an
independent test sample of size $n=500$ and then report the
averaged ME over $500$ realizations. The other measures are the
average model size (Size; defined as the number of nonzero
parameter estimates), the average number of false positives (FP;
defined as the number of nonzero estimates for zero parameters),
the average number of false negatives (FN; defined as the number
of zero estimates for nonzero parameters), and the proportion of
correct selections (C).

Table \ref{tbl-simu} indicates that MIC performs similarly to BSS
with BIC across all three models. In addition, all performance
measures of MIC improve as the sample size increases. By comparing
MIC against the other regularization methods, we find that MIC
outperforms them in general, except for the Gaussian linear
regression case where its performance is only comparable. We think
this is mainly because the objective function of MIC involves the
Gaussian profile likelihood $ n \ln \parallel \y - \X \bfbeta
\parallel^2$, which is nonconvex, while regularization methods can
work with the convex least squares problem $\parallel \y - \X
\bfbeta \parallel^2$ directly. Nevertheless, they all have to deal
with the same log-likelihood function in Model B and C. Note that
no implementation of MCP is available for the log-linear
regression, hence it is not presented for Model C. In sum, MIC not
only enjoys computational efficiency, but also demonstrates good
finite sample performance.

We next evaluates the standard error formula for nonzero parameter
estimates. Table \ref{tbl-se} presents the median absolute
deviation (MAD) value of $\widetilde{\bfbeta}_{(1)}$ out of 500
runs, which provides a more robust estimates of its standard
deviation. This MAD value matches reasonably well with the median
of standard errors of $\widetilde{\bfbeta}_{(1)}$. Also presented
is the MAD of standard errors.

Table \ref{tbl-gamma} presents the empirical size and power
results in testing $H_0:~ \gamma_j =0$ at the significance level
$\alpha=0.05$ over 1,000 simulation runs. It can be seen that the
proposed testing procedure has empirical sizes close to the
nominal level $0.05$ while showing excellent empirical power. This
result pertains closely to the super-efficiency phenomenon (see,
e.g., Chapter 8 of \citeauthor{van.der.Vaart.1998},
\citeyear{van.der.Vaart.1998}). Although super-efficiency could
occur on at most a Lebesgue null set, it does seem to have an
impact practically.

\renewcommand{\tabcolsep}{3.pt}
\renewcommand{\arraystretch}{1.2}
\renewcommand{\baselinestretch}{1}
\begin{table}
\begin{small}
\caption{Illustration of MIC with real data examples.}
\label{tbl-examples} \vspace{.02in}
\begin{center}
\centering
\begin{tabular}{lcccccccccccccc}
\multicolumn{15}{c}{(a) Linear Regression with Diabetes Data} \\ \hline
    &   \multicolumn{2}{c}{Full Model}          &&  \multicolumn{2}{c}{Best Subset}           &&  \multicolumn{5}{c}{MIC}    &&& \\ \cline{2-3} \cline{5-6} \cline{8-12}
    &   $\hat{\beta}_j$    &   SE  &&  $\hat{\beta}_j$    &   SE  && $\hat{\gamma}_j$ & SE & P-Value & $\hat{\beta}_j$    &   SE  &   LASSO   &   SCAD    &   MCP \\   \hline
\texttt{age} &   $-0.006$  &   0.037   &&      &       &&   0.000 & 0.036 & 1.000   &       &       &       &       \\
\texttt{sex} &   $-0.148$  &   0.038   &&  $-0.146$  &   0.037   && $-0.315$ & 0.037 & 0.000 &  $-0.137$  &   0.037   &   $-0.122$  &   $-0.149$  &   $-0.143$  \\
\texttt{bmi} &   0.321   &   0.041   &&  0.323   &   0.040   && 0.406 & 0.041 & 0.000 & 0.325   &   0.040   &   0.323   &   0.321   &   0.328   \\
\texttt{map} &   0.200   &   0.040   &&  0.202   &   0.039   && 0.344 & 0.040 & 0.000 & 0.196   &   0.039   &   0.184   &   0.199   &   0.197   \\
\texttt{tc}  &   $-0.489$  &   0.257   &&      &       &&  0.000 & 0.257 & 1.000    &       &   & $-0.064$  &   $-0.381$  &       \\
\texttt{ldl} &   0.294   &   0.209   &&      &       &&   0.000 & 0.209 & 1.000   &       &     &  &    0.216   &   $-0.067$  \\
\texttt{hdl} &   0.062   &   0.131   &&  $-0.179$  &   0.041   && $-0.332$ & 0.131 & 0.011 & $-0.171$  &   0.041   &   $-0.138$  &       &   $-0.179$  \\
\texttt{tch} &   0.109   &   0.100   &&      &       &&  0.000 & 0.099 & 1.000 &    &       &       &   0.080   &       \\
\texttt{ltg} &   0.464   &   0.106   &&  0.293   &   0.041   &&  0.390 & 0.106 & 0.000 & 0.294   &   0.041   &   0.318   &   0.426   &   0.300   \\
\texttt{glu} &   0.042   &   0.041   &&      &       &&  0.000 & 0.040 & 1.000 &    &       &   0.034   &   0.041   &   0.030   \\   \hline
BIC & \multicolumn{2}{c}{998.00} && \multicolumn{2}{c}{975.82} && \multicolumn{5}{c}{975.82} & 982.62 & 986.07 & 1001.64  \\ \hline
\end{tabular}

\vspace{.1in}
\begin{tabular}{lcccccccccccccc}
\multicolumn{15}{c}{(b) Logistic Regression with Heart Data} \\ \hline
    &   \multicolumn{2}{c}{Full Model}          &&  \multicolumn{2}{c}{Best Subset}           &&  \multicolumn{5}{c}{MIC}    &&& \\ \cline{2-3} \cline{5-6} \cline{8-12}
    &   $\hat{\beta}_j$     &   SE  &&  $\hat{\beta}_j$     &   SE  && $\hat{\gamma}_j$ & SE & P-Value &  $\hat{\beta}_j$     &   SE  &   LASSO   &   SCAD    &   MCP \\   \hline
\texttt{intercept}   &   $-0.845$  &   0.120   &&  $-0.847$  &   0.120   && $-0.842$ & 0.122 & 0.000 &  $-0.842$  &   0.119   &   $-0.787$  &   $-0.846$  &   $-0.844$  \\
\texttt{sbp} &   0.118   &   0.115   &&      &       &&  0.000 & 0.116 & 1.000 &    &       &   0.041   &       &   0.062   \\
\texttt{tobacco} &   0.365   &   0.120   &&  0.371   &   0.117   &&  0.418 & 0.123 & 0.001 & 0.349   &   0.116   &   0.299   &   0.371   &   0.369   \\
\texttt{ldl} &   0.383   &   0.119   &&  0.347   &   0.112   && 0.407 & 0.120 & 0.001 & 0.326   &   0.111   &   0.271   &   0.350   &   0.368   \\
\texttt{famhist} &   0.463   &   0.111   &&  0.456   &   0.110   && 0.476 & 0.112 & 0.000 & 0.446   &   0.110   &   0.371   &   0.456   &   0.460   \\
\texttt{obesity} &   $-0.146$ &   0.123   &&      &       &&  0.000 & 0.121 & 1.000    &       &       &   $-0.011$  &   $-0.086$  \\
\texttt{alcohol} &   0.015   &   0.109   &&      &       && 0.000 & 0.101 & 1.000 &     &       &       &       &       \\
\texttt{age} &   0.621   &   0.149   &&  0.643   &   0.142   && 0.656 & 0.152 & 0.000 & 0.656   &   0.142   &   0.544   &   0.645   &   0.632   \\   \hline
BIC & \multicolumn{2}{c}{532.26} && \multicolumn{2}{c}{516.12} && \multicolumn{5}{c}{516.12} & 526.14 & 521.11 & 521.44  \\ \hline
\end{tabular}

\vspace{.1in}
\begin{tabular}{lcccccccccccccccc}
\multicolumn{14}{c}{(c) Log-Linear Regression with Fish Data} \\
\hline
    &   \multicolumn{2}{c}{Full Model}          &&  \multicolumn{2}{c}{Best Subset}           &&  \multicolumn{5}{c}{MIC}    &&\\ \cline{2-3} \cline{5-6} \cline{8-12}
    &   $\hat{\beta}_j$     &   SE  &&  $\hat{\beta}_j$     &   SE  &&  $\hat{\gamma}_j$ & SE & P-Value & $\hat{\beta}_j$     &   SE  &   LASSO   &   SCAD    \\   \hline
\texttt{intercept}   &   $-0.360$  &   0.090   &&  $-0.313$  &   0.073   &&  $-0.395$ & 0.091 & 0.000 & $-0.304$  &   0.073   &   0.357   &   $-1.233$  \\
\texttt{nofish}  &   $-0.033$  &   0.059   &&      &       &&  0.000 & 0.061 & 1.000 &    &       &       &       \\
\texttt{livebait}    &   0.129   &   0.090   &&      &       && 0.000 & 0.081 & 1.000 &    &       &       &   0.426   \\
\texttt{camper}  &   $-0.010$  &   0.051   &&      &       &&  0.000 & 0.053 & 1.000 &    &       &       &       \\
\texttt{persons} &   0.047   &   0.057   &&      &       && 0.000 & 0.059 & 1.000 &     &       &       &       \\
\texttt{child}   &   $-0.653$  &   0.103   &&  $-0.643$  &   0.098   && $-0.639$ & 0.106 & 0.000 & $-0.638$  &   0.098   &       &   $-0.778$  \\
\texttt{xb}  &   1.447   &   0.064   &&  1.467   &   0.034   &&  1.464 & 0.067 & 0.000 & 1.464   &   0.034   &   0.331   &   1.012   \\
\texttt{zg}  &   0.659   &   0.136   &&  0.604   &   0.067   &&  0.606 & 0.142 & 0.000 & 0.604   &   0.067   &       &   0.283   \\
\texttt{xb:zg}   &   $-0.034$  &   0.059   &&      &       &&  0.000 & 0.061 & 1.000 &    &       &   0.176   &   $-0.001$  \\   \hline
BIC & \multicolumn{2}{c}{636.551} && \multicolumn{2}{c}{613.91} && \multicolumn{5}{c}{613.91} & 850.54 & 621.08   \\ \hline
\end{tabular}
\end{center}
\end{small}
\end{table}

\subsection{Real Data Examples}
\label{sec-data-example}

We consider the diabetes data \citep{Efron.etal.2004}, the heart
data \citep{Hastie.Tisshirani.Friedman.2009}, and the fish count
data (available at
\url{http://www.ats.ucla.edu/stat/data/fish.csv}) to illustrate
linear regression, logistic regression, and log-linear regression
models, respectively.

Table \ref{tbl-examples} shows that MIC (with $\lambda_0=\ln(n)$
and $a=10$) provides the similar selection as the best subset
selection across all three examples. In addition, the resulting
MIC estimates and their standard errors are quite close to these
of the BIC model. This finding indicates that MIC approximates the
best subset selection method well. This, together with MIC's
computational efficacy, allows us to employ MIC on data with large
numbers of covariates, even when BSS becomes infeasible. In the
diabetes data, it is particularly interesting to note that the
sign of the parameter estimate on \texttt{hdl} is positive under
the full model fitting, but becomes negative in MIC and several
other methods. This sign change could be problematic for
sign-constrained methods such as NG \citep{Breiman.1995}, but it
comes out naturally in MIC.

\begin{figure}[hp]
\centering
  \includegraphics[scale=0.7, angle=0]{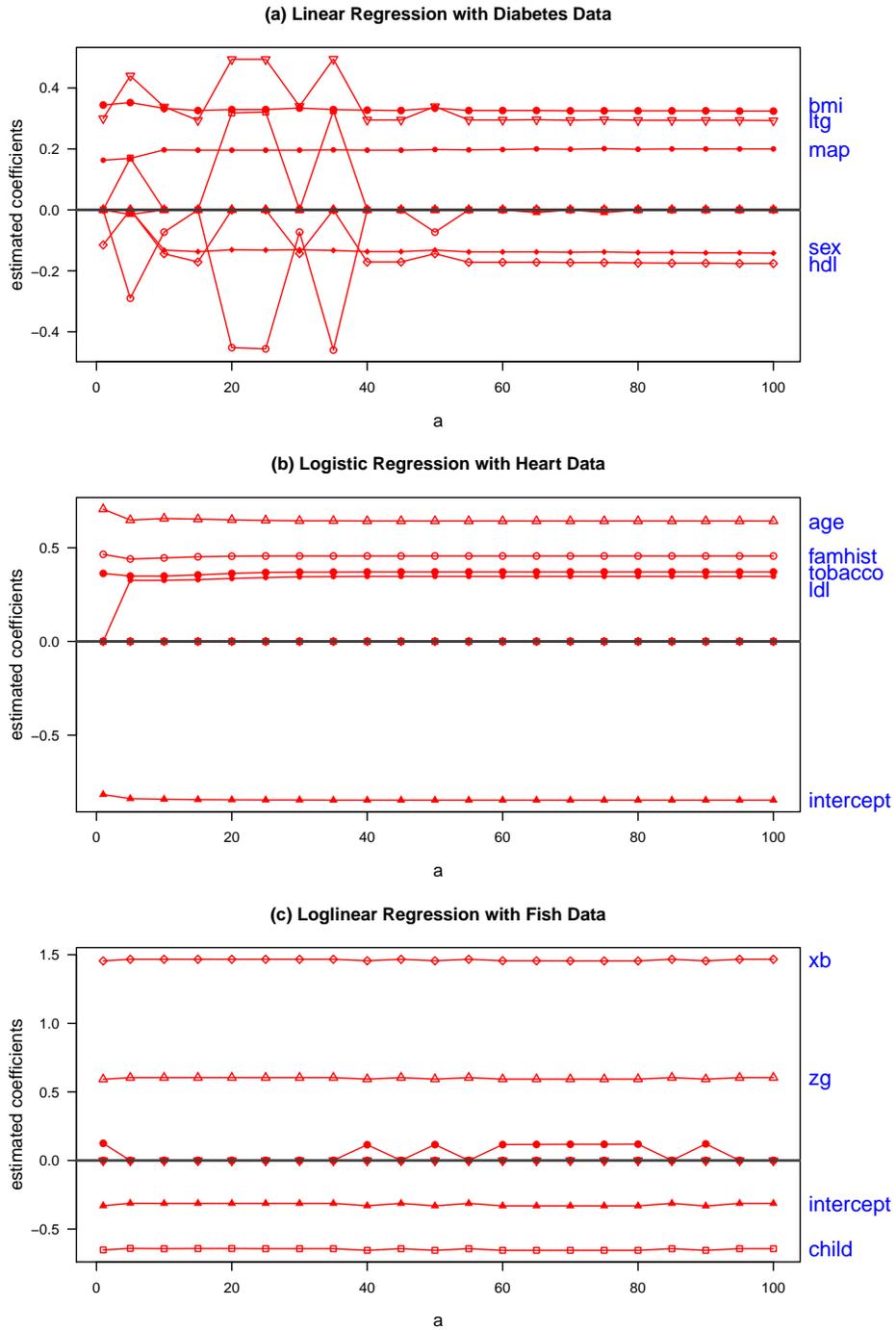}
  \caption{Illustrating the robustness of MIC with respect to the choice of $a$ in three real examples.
 The values of $a$ considered are $\{1, 5, 10, 15, \ldots, 100\}$.  \vspace{1in}
\label{fig3}}
\end{figure}

To illustrate the stability of MIC with respect to the value of
$a$, we obtain the MIC estimates for $a \in \{1, 5, 10, 15,
\ldots, 100\}$ and then plot them in Figure \ref{fig3}. While
there are some reasonable minor variations mainly owning to the
non-convex optimization nature, almost all the estimated
coefficients are quite steady in all three examples, showing that
the MIC estimation is generally robust to the choice of $a.$

\section{Discussion}
MIC is the first method that does sparse estimation by explicitly
approximating BIC. BIC is optimal in two aspects: it approximates
the posterior distribution of candidate models besides being
selection-consistent. This is why BIC has been used as an ultimate
yardstick in various variable selection and regularization
methods. MIC extends the best subset selection to scenario with
large $p$ by optimizing an approximated BIC. Formulated as a
smooth optimization problem, MIC is computationally advantageous
to the discrete-natured best subset selection and enjoys the
additional benefit in avoiding the post-selection inference.
Moreover, the search space in MIC remains to be the entire
parameter space. This explains why we expect MIC to outperform
many regularization methods that have a much reduced search space
for minimum BIC. By borrowing the knowledge of the fixed penalty
parameter for model complexity in BIC, MIC circumvents the tuning
parameter selection problem and hence is also computationally
advantageous to regularization methods.

Although the hyperbolic tangent function has been used to
approximate the cardinality in MIC, it can be replaced by other
unit dent functions. Since one focus of this paper is on the
variable selection consistency, we have adopted BIC by taking
$\lambda_0=\ln(n)$. In contrast, if the aim is on the model
selection efficiency or predictive accuracy, then we can adopt AIC
by setting $\lambda_0=2$. It can be shown that the resulting MIC
is selection-efficient by applying similar techniques to those
used in \citet{Zhang.etal.2010}. In sum, we can obtain variants of
MIC by changing its penalty function $w$ and penalty parameter
$\lambda_0$ to meet practical needs.

To broaden the usefulness of MIC, we conclude this article by
discussing three possible avenues for future research. First,
generalize MIC by accommodating the grouped or structured sparsity
(see, e.g., \citeauthor{Huang.Zhang.2010},
\citeyear{Huang.Zhang.2010}). Secondly, extend MIC to other
complex model or dependence structures, such as finite mixture
models, longitudinal data, and structural equation modelings
(SEM). Similar ideas may be applied to approximate the effective
degrees of freedom as well. In these settings, MIC can be
particularly useful because the log-likelihood function is not
concave and having convex penalties does not help anything with
the optimization problem. Thirdly, develop the MIC method for data
with diverging $p \rightarrow \infty$ yet $p/n \rightarrow 0$
\citep{Fan.Peng.2004} or ultra-high dimensions with $p \gg n$
\citep{Fan.Lv.2008} by approximating the extended or generalized
BIC as pioneered by \citet{Chen.Chen.2008}.

\vspace{.2in}

\newpage

\appendix
\begin{center}
{\Large APPENDIX: PROOFS}
\end{center}

\section{Proof of Theorem 1}
We first establish (i) by checking conditions in Theorem 1 of
\citet{Fan.Li.2001}. Note that the quantity $ p_{\lambda_n}
(|\beta_j|)$ corresponds to
$$p_{\lambda_n} (|\beta_j|) = \frac{\ln(n)}{2n} \cdot w(\gamma_j).$$
in MIC. Some quantities involved in the reparameterization $\beta
= \gamma w(\gamma)$ are summarized below:
\begin{equation*}
\left\{ \begin{array}{lclcl}
\dot{w} &=& d w(\gamma) / d \gamma &=& 2 a_n \gamma (1- w^2) \\
\ddot{w} &=& d^2 w(\gamma) / d \gamma^2 &=&  2a_n (1-w^2)(1-4a_n \gamma_{0j}^2 w) \\
w &=& \tanh(a_n \gamma^2) &=& \left(e^{a_n \gamma^2} - e^{-a_n \gamma^2}\right)/ \left( e^{a_n \gamma^2} + e^{-a_n \gamma^2} \right) \\
1-w^2 &=& \mbox{sech}(a_n \gamma^2) &=& 2 / \left( e^{a_n \gamma^2} + e^{-a_n \gamma^2} \right) \\
\end{array}
\right.
\end{equation*}
Since $a_n = O(n)$, $\gamma \rightarrow \beta$ and $w(\gamma)
\rightarrow 1$ for $\beta \neq 0$. It follows that, for $\beta
\neq 0$,
\begin{eqnarray*}
\dot{p}_{\lambda_n} (|\beta|) &=&  \frac{d p_{\lambda_n} (|\beta|)}{d \, \beta} \, = \,  \frac{\ln(n)}{2n} \, \frac{\dot{w}}{w + \gamma \dot{w}} \\
&=& \frac{\ln(n)}{n} \, \frac{a_n \gamma (1-w^2)}{w + 2 a_n \gamma^2 (1-w^2)} \\
&=& \frac{\ln(n)}{n} \, \frac{2 a_n \gamma}{e^{a_n \gamma^2} + e^{-a_n \gamma^2}  + 4 a_n \gamma^2} \\
&=& \frac{\ln(n)}{n} \, O\left\{ a_n e^{-a_n \gamma^2} \right\} \\
&=&  o( 1/ \sqrt{n}).
\end{eqnarray*}
Hence, $\max_j \left\{ \dot{p}_{\lambda_n} (|\beta_{0j}|):~
\beta_{0j} \neq 0 \right\} = o( 1/ \sqrt{n}).$ Similarly, it can
be shown that, for $\beta \neq 0$,
$$ \ddot{p}_{\lambda_n} (|\beta|) = \frac{d^2 p_{\lambda_n} (|\beta|)}{d \, \beta^2} = \frac{\ln(n)}{2n} \, \frac{ w \ddot{w} - 2 \dot{w}^2}{(w + \gamma \,
\dot{w})^3} ~ \stackrel{p}{\longrightarrow} ~ 0.$$ and so is
$\max_j \left\{ \ddot{p}_{\lambda_n} (|\beta_{0j}|):~ \beta_{0j}
\neq 0 \right\}$.

Therefore, there exists a local minimizer $\widetilde{\bfbeta}$ of
$Q_n(\bfbeta)$ such that $\parallel \widetilde{\bfbeta} -
\bfbeta_0 \parallel = O_p (1/\sqrt{n})$ by Theorem 1 of
\citet{Fan.Li.2001}. \hfill $\square$

\vspace{.1in}

\noindent To establish sparsity of $\widetilde{\bfbeta}_{(0)}$ in
(ii), it suffices to show that, for any $\sqrt{n}$-consistent
$\bfbeta = (\bfbeta_{(1)}^T, \bfbeta_{(0)}^T )^T$ such that
$\parallel \bfbeta_{(1)} - \bfbeta_{0(1)} \parallel = O_p
(1/\sqrt{n})$ and $\parallel \bfbeta_{(0)} \parallel =
O_p(1/\sqrt{n}),$ we have
\begin{equation}
\frac{\partial Q_n (\bfbeta)}{\partial \beta_j} = \begin{cases}
> 0 & \mbox{~if~} \beta_j >0 \\
< 0 & \mbox{~if~} \beta_j < 0
\end{cases}
\label{partial-deriv-Qn}
\end{equation}
for any component $\beta_j$ of $\bfbeta_{(0)}$ with probability
tending to 1 as $n \rightarrow \infty.$

Consider
$$
\frac{\partial Q_n(\bfbeta)}{\partial \beta_j} ~=~ - \,
\frac{2}{n}\,  \frac{\partial l(\bfbeta)}{
\partial \beta_j} + \frac{\ln(n)}{n} \cdot \frac{\partial w(\gamma_j)}{\partial \beta_j} \\
~=~ I + II$$ for $j = (q+1), \ldots, p$ when evaluated at
$\bfbeta.$ Note that $\beta_j = O_p(1/\sqrt{n})$ yet $\beta_j \neq
0$ for $\beta_j \in \bfbeta_{(0)}$. By standard arguments (see Fan
and Li, 2002) and using the fact that $\parallel \bfbeta -
\bfbeta_0 \parallel = O_p(1/\sqrt{n})$, it can be shown that the
first term $I$ is of order $O_p(1/\sqrt{n})$ under the regularity
conditions. For the second term $II,$ the analysis is more subtle,
depending on whether $a_n \gamma^2$ goes to 0, a constant, or
$\infty.$ Since it is desirable that
\begin{equation}
\label{w1-beta}
 \frac{\partial w(\gamma_j)}{\partial
\beta_j} = \frac{\dot{w}_j}{ w_j + \gamma_j \dot{w}_j} ~=~ \frac{
2 a \gamma_j (1-w_j^2)}{w_j + 2 a \gamma_j^2 (1 - w_j^2) } ~=~
\frac{4 a_n \gamma}{e^{a_n \gamma^2} + e^{-a_n \gamma^2}  + 4 a_n
\gamma^2}
\end{equation}
is $O_p (\sqrt{n})$ or even higher to have sparsity, neither the
choice $a_n \gamma^2=o(1)$ or $a_n \gamma^2 \rightarrow \infty$ is
not allowable because in either scenario, $\partial w(\gamma_j) /
\partial \beta_j$ is $o(1).$ Now set $a_n \gamma^2 = O_p(1)$. The
condition $\gamma w(\gamma) = \gamma \tanh(a_n \gamma^2) = \beta =
O_p(1/\sqrt{n})$ leads to the rate $\gamma = 1/\sqrt{n}$ and hence
$a_n = O(n).$ Therefore, the $O(n)$ rate for $a_n$ seems to be the
unique choice after taking all the side conditions into
consideration.

In this case, $\partial w(\gamma_j) / \partial \beta_j =
O_p(\sqrt{n})$. The second term becomes $II = O_p\left( \ln(n)
n^{1/2} / n \right) = O_p\left(\ln(n) /\sqrt{n} \right)$.
Moreover, it can be easily seen that the sign of $\partial
w(\gamma_j)/\partial \beta_j$ in (\ref{w1-beta}) is determined by
$\dot{w}_j$ and hence $\gamma_j$ or $\beta_j$, because $w_j \geq
0$ and $ \gamma_j \dot{w}_j \geq 0$. Put together, $\partial Q_n
(\bfbeta) / \partial \beta_j$ in (\ref{partial-deriv-Qn}) is
dominated by the second term $II$ and its sign is determined by
$\beta_j$. Therefore, the desired sparsity of
$\widetilde{\bfbeta}$ is established. \hfill $\square$

\vspace{.1in}

To show asymptotic normality of $\widetilde{\bfbeta}_{(1)}$ in
(ii), a close look at the proof of Theorem 2 in
\citet{Fan.Li.2001} reveals that it suffices to show that the
contribution from the penalty term to the estimating equation is
negligible relative to the gradient of the log-likelihood
function. More specifically, if we can show that
\begin{equation}
\frac{\ln(n)}{n} \, \left. \frac{\partial w(\gamma_j)}{\partial
\beta_j} \right|_{\beta_j = \tilde{\beta}_j} ~ = ~
o_p\left(\frac{1}{\sqrt{n}} \right), \label{A.3}
\end{equation}
for $j=1, \ldots , q,$ then Slutsky's theorem can be applied to
complete the proof. Equation  (\ref{A.3}) holds since, for any
non-zero $\beta_j \in \bfbeta_{0(1)}$, we have $\tilde{\beta}_j =
\beta_{j} + O_p( 1/\sqrt{n})$ and hence $\tilde{\gamma}_j =
\gamma_{j} + o_p(1)$ by the continuous mapping theorem, where
$\tilde{\beta}_j = \tilde{\gamma}_j w(\tilde{\gamma}_j)$ and
$\beta_{j} = \gamma_j w(\gamma_j).$ It follows that $\partial
w(\tilde{\gamma}_j) /
\partial \beta_j = o_p(1)$ in this case as shown earlier in the proof of (i).
Therefore $\dot{\rho}_n(\widetilde{\beta}_j) = o_p\{\ln(n)/n\} =
o_p(1/\sqrt{n}).$ The proof is completed. \hfill $\square$

\section{Proof of Theorem 2}

According to the definition, $\bfgamma_0$ is a constant that
depends on $n$ via $a_n$. In view of $\gamma- \beta = \gamma -
\gamma w(\gamma) = \gamma \{1- \tanh( a_n \, \gamma^2)\} = 2
\gamma / \{ \exp(2 a_n \gamma^2) +1)\},$ it follows that
$\left|\gamma_{0j} - \beta_{0j}\right| = O\{\exp(-2 a_n
\gamma_{0j}^2)\}$ for $\gamma_{0j} \neq 0$ and 0 otherwise. Hence
\begin{eqnarray*}
\parallel \bfgamma_0 - \bfbeta_0 \parallel_2 &\leq& \parallel
\bfgamma_0 - \bfbeta_0 \parallel_1 \, = \, \sum_{j=1}^q
\left|\gamma_{0j} - \beta_{0j}\right| \\
&\leq & \frac{2 q \max_{1 \leq j \leq q} \beta_j}{ \exp \{ 2 a_n
\, \min_{1 \leq j \leq q} \gamma_{0j}^2 \} +1 } \\
&=& O \left\{ \exp \{ - 2 a_n \, \min_{1 \leq j \leq q}
\gamma_{0j}^2 \} \right\}.
\end{eqnarray*}
Moreover, since the function $\beta= \gamma w(\gamma)$ is
continuous and so is its inverse, the continuous mapping theorem
yields $ \parallel \widetilde{\bfgamma} - \bfgamma_0 \parallel =
o_p(1).$

To study the asymptotic property of $\widetilde{\bfgamma},$ we
consider $\widetilde{\bfgamma}$ as a local minimizer of the
objective function $Q_n(\cdot)$, as stated in (\ref{SU00}). Since
$Q_n(\bfgamma)$ is smooth in $\bfgamma$, $\tilde{\gamma}$
satisfies the first-order necessary condition $\partial
Q_n(\widetilde{\bfgamma})/ \partial \bfgamma = \mathbf{0}$, which
gives
\begin{eqnarray}
&& - \frac{2}{n} \frac{\partial L(\widetilde{\bfbeta})}{\partial \bfbeta} \, \frac{\partial \widetilde{\bfbeta}}{\partial \bfgamma} + \frac{\ln(n)}{n} \frac{\partial \sum_j w(\tilde{\gamma}_j)}{\partial \bfgamma} = 0 \nonumber \\
& \Longrightarrow &  \nabla L(\widetilde{\bfbeta}) \, \diag\left(w_j + \tilde{\gamma}_j \dot{w}_j \right) = \frac{\ln(n)}{2} \left( \frac{d w_j}{d \gamma_j} \right)_{j=1}^p \nonumber \\
& \Longrightarrow & \nabla L(\widetilde{\bfbeta}) =
\frac{\ln(n)}{2} \left( \frac{\dot{w}_j}{w_j + \tilde{\gamma}_j
\dot{w}_j} \right)_{j=1}^p. \label{1st-order-bfgamma}
\end{eqnarray}
Next, applying Taylor's expansion of the LHS $ \nabla
L(\widetilde{\bfbeta})$ at $\bfgamma_0$ gives
$$ \frac{\ln(n)}{2} \left( \frac{\dot{w}_j}{w_j + \tilde{\gamma}_j \dot{w}_j} \right)_{j=1}^p = \nabla L(\bfbeta_0) + \nabla^2 L(\bfbeta_0) \, \left( \left.  \frac{\partial \bfbeta}{\partial \bfgamma} \right|_{\bfgamma = \bfgamma_0} \right)\,  (\widetilde{\bfgamma} - \bfgamma_0) + \mathbf{r}_n,$$
where $\mathbf{r}_n$ denotes the remainder term. It follows that
$$
\left( \left. \diag(w_j + \gamma_j \dot{w}_j) \right|_{\bfgamma =
\bfgamma_0} \right) \, (\widetilde{\bfgamma} - \bfgamma_0) \, = \,
\left\{ - \nabla^2 L(\bfbeta_0) \right\}^{-1} \, \left[ \nabla
L(\bfbeta_0)  - \frac{\ln(n)}{2} \left( \frac{\dot{w}_j}{w_j +
\tilde{\gamma}_j \dot{w}_j} \right)_{j=1}^p
 + \, \mathbf{r}_n \right].
$$
Therefore,
\begin{equation}
\sqrt{n} \left[ \mathbf{D}(\bfgamma_0) (\widetilde{\bfgamma} -
\bfgamma_0) + \mathbf{b}_n \right]  =  \left\{ - \frac{\nabla^2
L(\bfbeta_0)}{n} \right\}^{-1} \, \frac{\nabla
L(\bfbeta_0)}{\sqrt{n}}  + \mathbf{r}'_n, \label{eqn-A3}
\end{equation}
where $\mathbf{D}(\bfgamma_0)$ and $\mathbf{b}_n$ are defined in
(\ref{D-gamma0}) and (\ref{bias}), respectively, and the remainder
term is
$$ \mathbf{r}'_n = \left\{ - \frac{\nabla^2 L(\bfbeta_0)}{n} \right\}^{-1} \, \frac{\mathbf{r}_n}{\sqrt{n}}.$$
Under regularity conditions, standard arguments yield  $ \left\{ -
\nabla^2 L(\bfbeta_0) /n \right\}^{-1}
\stackrel{p}{\longrightarrow} \mathbf{I}^{-1}(\bfbeta_0)$; $\nabla
L(\bfbeta_0)/ \sqrt{n} \stackrel{d}{\longrightarrow}
\mathbf{N}\left\{\mathbf{0}, \, \mathbf{I}(\bfbeta_0) \right\};$
and $\mathbf{r}'_n = o_p(1)$ as $n \rightarrow \infty.$ Bringing
these results into (\ref{eqn-A3}) and an appeal to Slutsky's
Theorem give the desired asymptotic normality in
(\ref{asyp-norm-bfgamma}).

Note that the elements $D_{jj}$ of the diagonal matrix
$\mathbf{D}(\bfgamma_0)$ in (\ref{D-gamma0}) are evaluated at
$\bfgamma_0$. We have
$$D_{jj} = w(\gamma_{j0}) + \gamma_{j0} \, \dot{w}(\gamma_{j0}) =
\frac{e^{a_n \gamma_{j0}^2} - e^{-a_n \gamma_{j0}^2} - 4 a_n
\gamma_{j0}^2}{e^{a_n \gamma_{j0}^2} + e^{-a_n \gamma_{j0}^2}}.$$
Since $a_n=O(n)$, it can be seen that $\lim_{n \rightarrow \infty}
D_{jj}  = 1$ if $\gamma_{0j} \neq 0$ and $0$ otherwise.

To study the limit of bias $\mathbf{b}_n$, we rewrite (\ref{bias})
as
\begin{equation}
\mathbf{b}_n = \left\{ - \frac{\nabla^2 L(\bfbeta_0)}{n}
\right\}^{-1} \, \frac{\ln(n)}{2 \sqrt{n}} \left(
\frac{1}{\sqrt{n}} \, \frac{\dot{w}_j}{w_j + \tilde{\gamma}_j
\dot{w}_j} \right)_{j=1}^p. \label{bias1}
\end{equation}
Note that $\left\{ - \nabla^2 L(\bfbeta_0)/n \right\}^{-1}
\stackrel{p}{\longrightarrow} \mathbf{I}^{-1}(\bfbeta_0) \succ 0$
is evaluated at the constant $\bfbeta_0$ or $\bfgamma_0$ while the
last term of $\mathbf{b}_n$, with components $\dot{w}/\{\sqrt{n}
\, (w + \gamma \dot{w})\}$, is evaluated at
$\widetilde{\bfgamma}$. For $\gamma_{0j} \neq 0$, we have
$\tilde{\gamma}_j = \gamma_{0j} + o_p(1)$; for $\gamma_{0j} = 0$,
we have $\tilde{\gamma}_j = O_p(1/\sqrt{n}).$ Consider
\begin{equation}
\frac{\dot{w}}{w + \gamma \dot{w}} = \frac{4 a_n \gamma}{e^{a_n
\gamma^2} - e^{-a_n \gamma^2} + 4 a_n \gamma^2}.
\label{bias-term1}
\end{equation}
When $\tilde{\gamma}_j = \gamma_{0j} + o_p(1)$, $a_n
\tilde{\gamma}_j^2 \rightarrow \infty$ and hence
(\ref{bias-term1}) $= O_p(a_n\, e^{-a_n \gamma_{0j}^2}) = o_p(1);$
when $\tilde{\gamma}_j = O_p(1/\sqrt{n})$, we have shown
(\ref{bias-term1}) $= O_p( 1/\tilde{\gamma}_j) = O_p(\sqrt{n})$
earlier. Namely, the last term of $\mathbf{b}_n$ is $O_p(1)$ in
both cases. As a result, $\mathbf{b}_n = o_p(1)$ as $n \rightarrow
\infty$. Its componentwise convergence rates are exponential for
estimates of nonzero $\gamma_{0j}$'s and $O_p\{ \ln(n)/
\sqrt{n}\}$ for estimates of zero coefficients. This completes the
proof. \hfill $\square$



\end{document}